\documentclass[pra,twocolumn,showpacs]{revtex4}
\usepackage{epsfig}
\usepackage{amsbsy}
\usepackage{amsmath}
\usepackage{latexsym}

\input{tcilatex}

\begin{document}

\title{Decoherence in time evolution of bound entanglement}
\author{Zhe Sun}
\affiliation{Zhejiang Institute of Modern Physics, Department of Physics, Zhejiang
University, HangZhou 310027, China}
\author{Xiaoguang Wang}
\affiliation{Zhejiang Institute of Modern Physics, Department of Physics, Zhejiang
University, HangZhou 310027, China}
\author{Y. B. Gao}
\affiliation{Colleage of Applied Science, Beijing University of Techonology, Beijing,
100022, China}
\author{C. P. Sun}
\email{suncp@itp.ac.cn}
\homepage{www.itp.ac.cn/~suncp}
\affiliation{Institute of Theoretical Physics, Chinese Academy of Sciences, Beijing,
100080, China}
\date{\today}

\begin{abstract}
We study a dynamic process of disentanglement by considering the time
evolution of bound entanglement for a quantum open system, two qutrits
coupling to a common environment. Here, the initial quantum correlations of
the two qutrits are characterized by the bound entanglement. In order to
show the universality of the role of environment on bound entanglement, both
bosonic and spin environments are considered. We found that the bound
entanglement displays collapses and revivals, and it can be stable against
small temperature and time change. The thermal fluctuation effects on bound
entanglement are also considered.
\end{abstract}

\pacs{05.30.-d, 03.65.Ud, 75.10.Jm}
\maketitle

\section{Introduction}

Entanglement~\cite{Ein}, as an essential feature of quantum mechanics, helps
us to distinguish the classical and quantum nature of matter world. It has
become a key ingredient in quantum information processing, such as quantum
computing, quantum teleportation and quantum cryptography~\cite{Nielsen}-%
\cite{Ekert}. On the other hand, generally a realistic system is surrounded
by an environment. Thus the effects of the quantum decoherence such as
quantum dephasing on quantum entanglement should be considered for quantum
opene systems. It is reasonable that when we study quantum effects induced
by entanglement, the two-particle system should hold phase relations between
the components of the entangled states. Thus perceivably, due to the
interactions with environment, we can expect that the dephasing of
two-particle system can demonstrate some exotic properties.

Recently, Yu and Eberly~\cite{Yu} showed that two entangled qubits became
completely disentangled in a finite time under the influence of pure vacuum
noise. Surprisingly, they found that the behaviors of local decoherence is
different from the spontaneous disentanglement. The decoherence effects take
an infinite time evolution under the influence of vacuum noise while the
entanglement displays a ``sudden death" in a finite time. In their
investigations and other studies on disentanglement in open quantum systems~%
\cite{Zubairy}~\cite{Roszak}, only qubit systems are considered. Here, the
disentanglement process is characterized by time evolution of the
concurrence~\cite{Conc}. It is well-known that the concurrence is not
available in the higher-dimensional systems.

For the systems with spins larger than $1/2$, one can use the positive
partial transpose (PPT) method~\cite{Horodecki1} to study disentanglement.
For the mixed states of two spin halves and (1/2 ,1) mixed spins, the PPT
method can fully characterize entanglement. However, in the case of two
qutrits and even larger spins, one only know that if a state does not have
PPT, the state must be entangled. In other words, one may use the method to
witness entanglement. Actually, in the case of higher dimension, there are
two qualitatively different types of entanglement~\cite{Horodecki2}, free
entanglement (FE) which corresponds to the states without a PPT, and bound
entanglement (BE) corresponding to the entangled states, however, with a
PPT. The BE is an intrinsic property and cannot be distilled to a singlet
form, thus it cannot be used alone for quantum communication. Nevertheless,
the BE can be activated and then contribute to quantum communication~\cite%
{Horodecki3}. A formal entanglement-energy analogy~\cite{Vedral} implies
that the bound entanglement is like the energy of a system confined in a
shallow potential well. If we add a small amount of extra energy, behaving
as a perturbation, to the system, its energy can be deliberated. The
existence of bound entangled states reveals a transparent form of
irreversibility in entanglement processing~\cite{Horodecki4}.

In this paper, we consider an open composite system, a two-qutrit system
commonly coupled to an environment, and study a type of dynamical process of
disentanglement, where the two qutrits are initially prepared in a bound
entangled state. We would like to reveal that different environment gives
different dynamics of entanglement. Firstly, a bosonic heat bath is
considered. We remark that this modeling of environment is universal~\cite%
{Leggett,sun} in the sense that any environment weakly coupled to a system
can be approximated by a collection of harmonic oscillators. Secondly, we
consider a spin environment consisting of spin halves which can be
considered as a fermionic environment. We let two types of bound entangled
states being initial state of the two qutrits in order to find the different
properties of bound entangled state during the quantum dephasing. And
initially the environments are assumed to be at thermal equilibrium states,
which helps us to find effects of the thermal fluctuation on dynamics of
quantum entanglement.

This paper is organized as follows. In Sec.~II, we consider the bosonic
environment and give the analytical results of FE and BE. we numerically
study the BE to illustrate the details of the dynamics of entanglement. In
Sec.~III, the two qutrits are coupled to a spin environment. Also the
analytical and numerical results are given to show the effects of coupling
strength, temperature and energy spectrum structure on the dynamical
behaviors of entanglement. The conclusion is given in Sec. IV.

\section{Bound entanglement in a bosonic environment}

We start with a well known model of the pure dephasing~\cite{sun,gao}, where
two qutrits interact with the environment, which is modelled as a heat bath
with many harmonic oscillators of frequency $\omega _{j}$. The model
Hamiltonian reads
\begin{equation}
H=\sum_{j}^{L}H_{j}=\sum_{j}^{L}[\hbar \omega _{j}b_{j}^{\dag
}b_{j}+g(b_{j}^{\dag }+b_{j})(S_{1z}+S_{2z})],  \label{1}
\end{equation}%
where $b_{j}^{\dag }$ and $b_{j}$ are creation and annihilation operators,
respectively, $S_{1z}$ and $S_{2z}$ are $z$ components of two spin-1
operators, and $g$ denotes the coupling strength between the spins and the
heat bath.

In order to study the dynamical process of entanglement in our
system, it is convenient for us to study the time evolution in the
interaction picture. Here,
\begin{equation}
H_{0}=\sum_{j}\hbar \omega _{j}b_{j}^{\dag }b_{j}
\end{equation}%
is the free Hamiltonian, and the interaction Hamiltonian
\begin{equation}
H_{I}=\sum_{j}g(b_{j}^{\dag }+b_{j})(S_{1z}+S_{2z}).
\end{equation}%
Then, through the Wei-Norman method~\cite{Wei-Norman}, the time evolution
operator in the interaction picture is factorized as,
\begin{equation}
U\left( t\right) =\prod\limits_{j}e^{i\Phi _{j}(t)S_{z}^{2}}D\left[
z_{j}(t)S_{z}\right] ,  \label{u}
\end{equation}%
where $S_{z}=S_{1z}+S_{2z},$
\begin{eqnarray}
\Phi _{j}(t) &=&\frac{g^{2}}{\hbar ^{2}\omega _{j}^{2}}\left( \omega
_{j}t-\sin \omega _{j}t\right) , \\
z_{j}(t) &=&\frac{g}{\hbar \omega _{j}}\left( 1-e^{i\omega _{j}t}\right)
\end{eqnarray}%
and $D(z_{j}S_{z})=\exp \left[ {\left( z_{j}b_{j}^{\dag }-z_{j}^{\ast
}b_{j}\right) S_{Z}}\right] $ is the displacement operator.

Before discussing the dynamical process of entanglement, we introduce two
quantities to quantitatively study entanglement. One is the negativity~\cite%
{Vidal}, which can be used to study FE. For a state $\rho$, negativity is
defined in terms of the trace norm of the partial transposed matrix
\begin{equation}
\mathcal{N(\rho )}=\frac{\Vert \rho ^{T_{1}}\Vert _{1}-1}{2},
\end{equation}
where $T_{1}$ denotes the partial transpose with respect to the first
subsystem. If $\mathcal{N}>0$, then the two-spin state is free entangled. As
an entanglement measure, the negativity is operational and easy to compute,
and it has been used to characterize entanglement in large spin system very
well~\cite{Schliemann}-~\cite{Zhe}.

In order to characterize BE, one can use the so-called realignment criterion
(cross-norm criterion) which proved to be very efficient~\cite{Realign}. The
operation of realignment on the density matrix is just as $(\rho
^{R})_{ij,kl}=\rho _{ik,jl}$. A separate state $\rho $ always satisfies $%
||\rho ^{R}||\leq 1.$ Thus, a quantity for the BE can be defined as
\begin{equation}
\mathcal{R(}\rho )=\max \left\{ 0,||\rho ^{R}||-1\right\} .
\end{equation}
We call this the witness quantity. Only when $\mathcal{R(}\rho )>0$ and $%
\mathcal{N(\rho )=}0$, the state is bound entangled.

\subsection{Horodecki's Bound entangled state}

In the following discussions, we consider the dynamical evolution process of
the two-spin 1 system, deriven by the Hamiltonian (\ref{1}) with the initial
state being in the Horodecki's bound entangled state~\cite{Horodecki3}.

\subsubsection{Analytical results}

The bound entangled state reads~\cite{Horodecki3}:
\begin{eqnarray}
\rho _{a}(0) &=&\frac{2}{7}\mathbf{P}_{+}+\frac{a}{7}\varrho _{+}+\frac{5-a}{%
7}\varrho _{-},  \label{BE1} \\
\ 2 &\leq &a\leq 5,  \notag
\end{eqnarray}%
where
\begin{eqnarray}
\mathbf{P}_{+} &=&|\Psi _{+}\rangle \langle \Psi _{+}|,|\Psi _{+}\rangle =
\notag \\
&&\frac{1}{\sqrt{3}}(|00\rangle +|11\rangle +|22\rangle ), \\
\varrho _{+} &=&\frac{1}{3}(|01\rangle \langle 01|+|12\rangle \langle
12|+|20\rangle \langle 20|),  \notag \\
\varrho _{-} &=&\frac{1}{3}(|10\rangle \langle 10|+|21\rangle \langle
21|+|02\rangle \langle 02|).
\end{eqnarray}%
where $\left\vert m_{1}m_{2}\right\rangle ,(m_{1},m_{2}=0,1,2)$\ are the
eigenvectors of $S_{z}=S_{1z}+S_{2z}$, with the corresponding eigenvalues $%
m_{1}+m_{2}-2$, respectively.

In Ref.~\cite{Horodecki2}, Horodecki demonstrated that
\begin{equation}
\rho _{a}\text{ is }\left\{
\begin{array}{l}
\text{separable for }2\leq a\leq 3, \\
\text{bound entangled for }3<a\leq 4, \\
\text{free entangled for }4<a\leq 5.%
\end{array}%
\right.
\end{equation}%
And the density matrix for the initial state\ of the total system\ is a
simple direct\ product
\begin{equation}
\rho _{\text{tot}}\left( 0\right) =\rho _{a}\otimes \rho _{E},  \label{ini}
\end{equation}%
where $\rho _{E}$ is the density matrix of environment.

Driven by the time evolution operator~(\ref{u}), the system will evolve from
the bound entangled state $\rho _{a}$ into the state described by
\begin{eqnarray}
\rho _{1,2}\left( t\right) &=&\text{Tr}_{E}\left[ U\left( t\right) \rho _{%
\text{tot}}\left( 0\right) U^{\dag }\left( t\right) \right]  \notag \\
&=&\frac{2}{21}\big[\left( \left\vert 00\right\rangle \left\langle
00\right\vert +\left\vert 11\right\rangle \left\langle 11\right\vert
+\left\vert 22\right\rangle \left\langle 22\right\vert \right)  \notag \\
&&+\left( F_{1}\left( t\right) \left\vert 00\right\rangle \left\langle
11\right\vert +\text{H.c.}\right)  \notag \\
&&+\left( F_{2}\left( t\right) \left\vert 22\right\rangle \left\langle
11\right\vert +\text{H.c.}\right)  \notag \\
&&+\left( F_{3}\left( t\right) \left\vert 00\right\rangle \left\langle
22\right\vert +\text{H.c}\right) \big]  \notag \\
&&+\frac{a}{7}\varrho _{+}+\frac{5-a}{7}\varrho _{-}.  \label{rho12}
\end{eqnarray}%
where
\begin{eqnarray}
F_{1}\left( t\right) &=&\text{Tr}_{E}\left[ \rho _{E}U_{1}^{\dag }\left(
t\right) U_{0}\left( t\right) \right]  \notag  \label{dec} \\
F_{2}\left( t\right) &=&\text{Tr}_{E}\left[ \rho _{E}U_{1}^{\dag }\left(
t\right) U_{2}\left( t\right) \right]  \notag \\
F_{3}\left( t\right) &=&\text{Tr}_{E}\left[ \rho _{E}U_{2}^{\dag }\left(
t\right) U_{0}\left( t\right) \right]
\end{eqnarray}%
are \emph{decoherence factors}\cite{sun}. The unitary operators $U_{0}\left(
t\right) $, $U_{1}\left( t\right) $, and $U_{2}\left( t\right) $ are derived
from Eq.~(\ref{u}) just by replacing operator $S_{z}=S_{1z}+S_{2z}$ with
numbers $-2,0$ and $2$, respectively.

From the reduced density matrix (\ref{rho12}), the realigned matrix becomes
\begin{eqnarray}
(\rho _{12}(t))^{R}&=&\frac{1}{21}\left( A_{_{3\times 3}}\oplus B_{2\times
2}^{(1)}\oplus B_{2\times 2}^{(2)}\oplus B_{2\times 2}^{(3)}\right) ,  \notag
\\
A_{_{3\times 3}}&=&\left(
\begin{array}{ccc}
2 & a & 5-a \notag \\
5-a & 2 & a \notag \\
a & 5-a & 2%
\end{array}%
\right) ,  \notag \\
B_{2\times 2}^{(k)} &=&\left(
\begin{array}{cc}
2F_{k} & 0 \\
0 & 2F_{k}^{\ast }%
\end{array}%
\right) (k=1,2,3).
\end{eqnarray}%
Then, the witness quantity $\mathcal{R}$ is obtained as \textit{\ }
\begin{eqnarray}
\mathcal{R(}\rho _{1,2}) &=&\max \left\{ ||\rho _{1,2}^{R}\left( t\right)
||-1,0\right\}  \notag \\
&=&\frac{2}{21}\max \big \{\sqrt{3a^{2}-15a+19}  \notag \\
&&+2\left( \left\vert F_{1}\right\vert +\left\vert F_{2}\right\vert
+\left\vert F_{3}\right\vert \right) -7,0\big \}.  \label{rrr}
\end{eqnarray}%
As mentioned above, the positive witness quantity can quantify the
nontrivial BE only when the negativity vanishes. Thus, we need to calculate
the time evolution of negativity.

We first make the partial transpose of $\rho_{12}$ with respect to the
second system and obtain
\begin{eqnarray}
(\rho _{12}(t))^{T_{2}} &=&\frac{1}{21}\left( C_{_{3\times 3}}\oplus
D_{2\times 2}^{(1)}\oplus D_{2\times 2}^{(2)}\oplus D_{2\times
2}^{(3)}\right) ,  \notag \\
C_{_{3\times 3}} &=&\left(
\begin{array}{ccc}
2 & 0 & 0 \notag \\
0 & 2 & 0 \notag \\
0 & 0 & 2%
\end{array}%
\right) ,  \notag \\
D_{2\times 2}^{(k)} &=&\left(
\begin{array}{cc}
a & 2F_{k} \\
2F_{k}^{\ast } & 5-a%
\end{array}%
\right) (k=1,2,3).
\end{eqnarray}%
Then, from the above equation, we immediately obtain the negativity
\begin{equation}
\mathcal{N(}\rho _{1,2})=\frac{1}{42}\sum\limits_{k=1}^{3}\max \big \{0,%
\sqrt{\left( 2a-5\right) ^{2}+16\left\vert F_{k}\right\vert ^{2}}-5\big \}.
\label{negativity}
\end{equation}%
Thus, we have obtained analytical expressions of witness quantity $\mathcal{R%
}$ and negativity $\mathcal{N}$ in terms of the three decoherence
factors. It is natural to see that if the decoherence factors are
zero, namely, the completely decoherence occurs, from
Eqs.~(\ref{rrr}) and (\ref{negativity}), we have
$\mathcal{R}=\mathcal{N=}0.$ From Eq.~(\ref{negativity}), we can
also see that in the region $3<a\leq 4$, negativity always\ gives
zero at any time since $\left\vert F_{k}\right\vert \leq 1$.

From the above discussions, once we know the decoherence factors, the two
quantities $\mathcal{R}$ and $\mathcal{N}$ for detecting entanglement can be
determined. So, we are left to obtain these decoherence factors. It is well
known that high temperature may enhance the decoherence, thus it is
reasonable\ to choose\ a thermal equilibrium state as the initial state of
the heat bath, which is described by the density matrix
\begin{eqnarray}
\rho _{E} &=&\prod\limits_{j}\rho _{Ej}=\prod\limits_{j}\frac{e^{-\beta
\hbar \omega _{j}b_{j}^{\dag }b_{j}}}{\text{Tr}\left[ e^{-\beta \hbar \omega
_{j}b_{j}^{\dag }b_{j}}\right] }  \notag \\
&=&\prod\limits_{j}\left( 1-e^{-\beta \hbar \omega _{j}}\right) e^{-\beta
\hbar \omega _{j}b_{j}^{\dagger }b_{j}},  \label{rho}
\end{eqnarray}%
where $\beta =1/k_{B}T$, $k_{B}$ is the Boltzmann's constant, and we choose $%
k_{B}=1$ for simplicity in the following.

For the bosonic environment we calculate the decoherence factors in the
coherent-state representation. The $P$-representation for the thermal state
is given by%
\begin{eqnarray}
\rho _{E}\left( 0\right) &=&\prod\limits_{j}\rho _{Ej}=\prod\limits_{j}\int
\rho _{j}\left( \alpha \right) \left\vert \alpha \right\rangle \left\langle
\alpha \right\vert d^{2}\alpha , \\
\rho _{j}\left( \alpha \right) &=&\frac{1}{\pi \left\langle
n_{j}\right\rangle }\exp \left( -\frac{\left\vert \alpha \right\vert ^{2}}{%
\left\langle n_{j}\right\rangle }\right) ,
\end{eqnarray}%
where $\left\langle n_{j}\right\rangle =\left( e^{\beta \hbar \omega
_{j}}-1\right) ^{-1}$\ is the thermal excitation number of harmonic
oscillators. From Eq.~(\ref{dec}) and using the $P$-representation,
one obtains the modulus of the decoherence factors \cite{C.P.Sun2}
\begin{eqnarray}
|F_{1}\left( t\right) | &=&\left\vert F_{2}\left( t\right) \right\vert
=\prod_{j}|\text{Tr}_{E_{j}}[\rho _{E_{j}}D\left( 2z_{j}\right) ]e^{i4\Phi
_{j}}|  \notag \\
&=&\prod_{j}e^{-2\left\vert z_{j}\right\vert ^{2}\left( 2\left\langle
n_{j}\right\rangle +1\right) }  \notag \\
&=&\exp (\sum_{j}\frac{-8g^{2}}{\hbar ^{2}\omega _{j}^{2}}\left(
2\left\langle n_{j}\right\rangle +1\right) \sin ^{2}\left( \frac{\omega _{j}t%
}{2}\right) , \\
|F_{3}\left( t\right) | &=&\prod_{j}\text{Tr}_{j}\left( \rho _{Bj}D\left(
-4z_{j}\right) \right) =|F_{1}\left( t\right) |^{4}.
\end{eqnarray}

As expected, the above three quantities\ are smaller than or equal to unity.
Now, we study the decoherence of BE, and choose parameter $a=4$ in the bound
entangled state in the following discussions. This choice of parameter
maximize the quantity $\mathcal{R}$. Then, Eq.~(\ref{rrr}) simplifies to
\begin{equation}
\mathcal{R(}\rho _{1,2})=\frac{2}{21}\max \left\{ 0,2\left( 2\left\vert
F_{1}\right\vert +\left\vert F_{1}\right\vert ^{4}\right) +\sqrt{7}%
-7\right\} .  \label{realignment}
\end{equation}%
Then, we find that the dynamic properties of BE is thus directly
related to the one single decoherence factors $|F_{1}(t)|$. By
numerical calculation, one obtains the threshold point of
\begin{equation}
\left\vert F_{1}\right\vert \approx 0.839829,  \label{fff}
\end{equation}%
before which the quantity $\mathcal{R}$ is larger than zero, implying that
the state is a bound entangled state. In Fig.~1, we numerically show the
modulus $|F_1(t)|$ versus time. The frequencies $\omega_j$ are chosen
randomly in a region $\omega_j\in[50, 55]$. Then, the modulus $|F_1(t)|$
oscillates with time and periodically crosses the horizontal line
corresponding to the threshold value $F_1=0.8398$. Obviously, the witness
quantity displays discontinuous behavior and below the line it becomes zero.

\begin{figure}[tbp]
\includegraphics[bb=28 155 572 600, width=5 cm, clip]{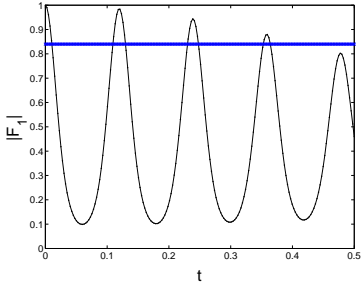}
\caption{The modulus of the decoherence factor $|F_1(t)|$ versus time with $%
g=2$. The frequencies of heat bath are chosen randomly in a region $\protect%
\omega_j\in[50, 55]$. The size of the bath is $L=200$ and the system
is at the temperature $T=1$. The horizontal line in the figure
corresponds to the threshold value $|F_1|=0.8398$.}
\end{figure}

In the following we consider some special cases of the energy distribution
in the environment and find that the decoherence factors decay as a Gaussian
or a exponential form with time, and consequently we know the time behaviors
of the BE.

$i)$ Let us\ choose a certain type of distribution of $\omega _{j}$ in the
region $\left[ 0,\omega \right] ,$ where $\omega $ is an arbitrary value
larger than zero. We do not care about the exact form of the distribution,
however it can be achieved for us to choose a sufficiently small cutoff
frequency $\omega _{j_{c}}$ to make sure that at a finite time, the
decoherence factor ($\hbar =1$)
\begin{eqnarray}  \label{bath1}
|F_{1}\left( t\right) | &\leq &\exp \sum_{j}^{j_{c}}\left[ \frac{-8g^{2}}{%
\omega _{j}^{2}}\sin ^{2}\left( \frac{\omega _{j}t}{2}\right) \left(
2\left\langle n_{j}\right\rangle +1\right) \right]  \notag \\
&\approx &\exp \left[ -2g^{2}\sum_{j}^{j_{c}}\left( 2\left\langle
n_{j}\right\rangle +1\right) t^{2}\right] =e^{-\gamma t^{2}},  \notag \\
\end{eqnarray}%
where
\begin{equation}
\gamma =2g^{2}\sum_{j}^{j_{c}}\left( 2\left\langle n_{j}\right\rangle
+1\right) .
\end{equation}%
It can be seen that the decoherence factor displays a Gaussian decay with
time. Moreover one may observe that the decay parameter $\gamma $ increases
at high temperature since $\left\langle n_{j}\right\rangle $\ is a
monotonically increasing function of temperature $T$, and enlarging the
strength $g$ can also increase $\gamma $. Substituting Eq.~(\ref{bath1}) to (%
\ref{realignment}) leads to%
\begin{equation}
\mathcal{R(}\rho _{1,2})=\frac{2}{21}\max \left\{ 0,2\left( 2e^{-\gamma
t^{2}}+e^{-4\gamma t^{2}}\right) +\sqrt{7}-7\right\},
\end{equation}%
and it will decay to zero in a fixed time $t_{0}$, which can be determined
from Eq.~(\ref{fff})
\begin{equation}
t_{0}=0.4176/\sqrt{\gamma }.
\end{equation}
When the evolution time is larger than the threshold value $t_0$, the BE
suddenly vanishes.

$ii)$ If we choose some continuous spectrum, the sum in the decoherence
factors (we assume$\ \hbar =1$) becomes%
\begin{equation}
\ln |F_{1}\left( t\right) |=-\sum_{j}\left[ \frac{8g_{j}^{2}}{\omega _{j}^{2}%
}\sin ^{2}\left( \frac{\omega _{j}t}{2}\right) \right] ,
\end{equation}%
where $g_{j}=g\sqrt{\left( 2\left\langle n_{j}\right\rangle +1\right) }.$
Assume a spectrum distribution $\rho \left( \omega _{j}\right) ,$ the above
equation becomes
\begin{equation}
\ln |F_{1}\left( t\right) |=-\int_{0}^{\infty }\frac{8\rho \left( \omega
_{j}\right) g_{j}^{2}}{\omega _{j}^{2}}\sin ^{2}\frac{\omega _{j}t}{2}%
d\omega _{j}.
\end{equation}%
For some concrete spectrum distributions, interesting circumstances may
arise. For instance, when $\rho \left( \omega _{j}\right) =\gamma /(2\pi
g_{j}^{2})$\ the integral converges to a negative number proportional to
time $t$, precisely, $|F_{1}\left( t\right) |=\exp({-\gamma t}),$ $%
|F_{3}\left( t\right) |=\exp({-4\gamma t})$. Thus, in this case, the
reasonable assumption on the energy distribution brings us a exponential
decay of decoherence factor and entanglement with time.

$iii)$ Now we will choose another more general distribution $\rho \left(
\omega \right) $. Assume that all the coefficients\ $g_{j}$\ are equal: $%
g_{j}=G$. If the frequencies lie within an interval \ $\left[ \omega
_{1},\omega _{2}\right] $\ and the distribution is homogeneous, we have $\
\rho \left( \omega _{k}\right) =N/\left( \omega _{2}-\omega _{1}\right) ,\ $%
thus \cite{C.P.Sun}
\begin{eqnarray}
&&\ln |F_{1}\left( t\right) |  \notag \\
&=&-\sum_{j}\left[ \frac{8g_{j}^{2}}{\omega _{j}^{2}}\sin ^{2}\frac{\omega
_{j}t}{2}\right]  \notag \\
&=&-\int_{\frac{\omega _{1}}{2}}^{\frac{\omega _{2}}{2}}\sin ^{2}\left(
\omega _{k}t\right) \frac{2G^{2}\rho \left( \omega _{k}\right) }{\omega
_{k}^{2}}d\omega _{k}  \notag \\
&=&\frac{-2G^{2}N}{\omega _{2}-\omega _{1}}\int_{\frac{\omega _{1}}{2}}^{%
\frac{\omega _{2}}{2}}\frac{\sin ^{2}\omega _{k}t}{\omega _{k}^{2}}d\omega
_{k}  \notag \\
&\leq &\frac{-2G^{2}N}{\omega _{2}-\omega _{1}}\frac{4}{\omega _{2}^{2}}%
\int_{\frac{\omega _{1}}{2}}^{\frac{\omega _{2}}{2}}\sin ^{2}\left( \omega
_{k}t\right) d\omega _{k}  \notag \\
&=&\frac{-2G^{2}N}{\omega _{2}^{2}}\left[ 1-\frac{2\cos \left( \frac{\omega
_{2}+\omega _{1}}{2}t\right) \sin \left( \frac{\omega _{2}-\omega _{1}}{2}%
t\right) }{\left( \omega _{2}-\omega _{1}\right) t}\right] .
\end{eqnarray}%
By substituting the above equation into the Eq.~(\ref{realignment}), we see
that when the environment has sufficiently large size $L$, the decoherence
factor and the quantity $\mathcal{R}$ will decay with time rapidly.

\subsubsection{Numerical results}

Next, we resort to numerical calculation to test the above analysis and show
more dynamic behaviors of entanglement.
\begin{figure}[tbp]
\includegraphics[bb=81 249 495 581, width=6 cm, clip]{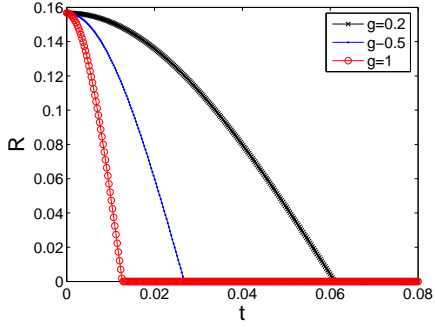}
\caption{$\mathcal{R}$ versus time with different coupling parameter $g$.
The frequencies of heat bath are chosen randomly in a low region $\protect%
\omega_j\in[0,5]$. The size of bath $L=200$ and the system is at a finite
temperature $T=1$.}
\end{figure}

\begin{figure}[tbp]
\includegraphics[bb=17 326 566 597, width=8 cm, clip]{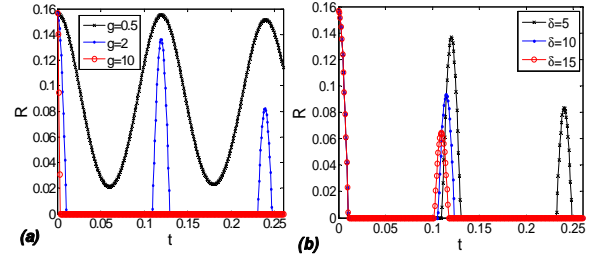}
\caption{\textbf{(a)} $\mathcal{R}$ versus time with different
coupling parameter $g$. The frequencies of heat bath are chosen
randomly in a higher region $\protect\omega_j\in[50, 55]$. The size
of the bath $L=200$ and the system is at the temperature $T=1$.
\textbf{(b)} We consider the frequency distribution randomly in
$[50,50+\protect\delta]$, the coupling $g=2$ and the system is at
$T=1$.}
\end{figure}

\begin{figure}[tbp]
\includegraphics[bb=13 200 585 661, width=8 cm, clip]{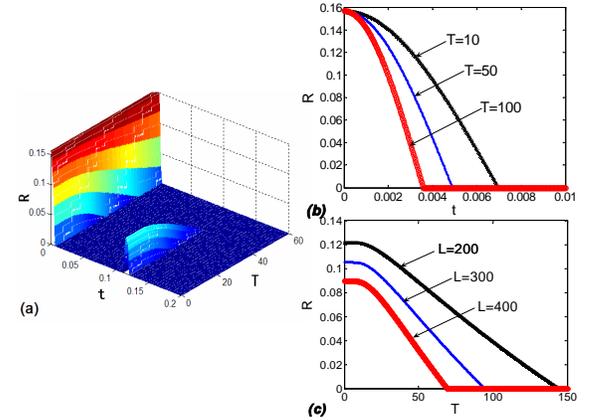}
\caption{\textbf{(a)} Three dimensional (3D) diagram of $\mathcal{R}$ versus
time and temperature, with $L=200$ and $g=3$. The region of frequencies is $%
\protect\omega _{j}\in \lbrack 50,58]$. \textbf{(b)} Quantity $\mathcal{R}$
versus time at different temperatures. \textbf{(c)} $\mathcal{R}$ versus
temperature at a fixed time $t=0.003$ for different sizes.}
\end{figure}
In Fig.~2, we choose a random distribution of the environment energy $\omega
_{j}$ over a finite region $\lbrack 0,\omega ]$, and give the time behaviors
of the quantity $\mathcal{R}$. A Gaussian decay is exhibited, and this is
consistent with the previous analysis (Eq.~(\ref{bath1})). Of course, larger
$g$ accelerates the decay process of the BE. When we change other
parameters, such as the width of frequencies and the temperature, the
quantity $\mathcal{R}$ displays similar behaviors as long as the region of $%
\omega _{j}$ begins from zero. This implies that the harmonic oscillators
with lower energies in the heat bath determine the behaviors of the BE. .

Instead of choosing the low energies of the harmonic oscillators in
environment, we consider a random distribution of $\omega _j$ in a higher
frequency region $[50,50+\delta]$, where $\delta$ is the width of the
distribution. The numerical results are shown in Fig.~3. For small values of
the coupling constants, as shown in Fig.~3 (a), the BE displays oscillations
with time. For $g=2$, we observe collapses and revivals of the BE. The
revivals result from the revivals of the decoherence factors. When the
coupling strength is strong enough, the BE decays rapidly to zero without
revivals.

Fig.~3(b) presents numerical results for different frequency widths.
Consider an extreme case $\delta=0$, that is, only one frequency is
taken into account. In this case, the BE displays periodic collapse
and revivals for large $g$, and the revival amplitude of
$\mathcal{R}$ is one. From the figure, we see that when the
frequency width increases, the revival amplitude decreases.
Increasing the width $\delta $ means that the harmonic oscillators
in the heat bath own much more different frequencies and the
entanglement revival will be suppressed.

Now we consider the thermal effect on the BE in Fig.~4. The
subfigures {(a)} and {(b)} show that the thermal fluctuation can
destroy entanglement and accelerate the decaying process. With the
joint effect of thermal fluctuation and strong coupling $g$,
entanglement vanishes in a finite time without reviving. In
Fig.~4{(c)}, we can see that with the temperature increasing the
entanglement decreases to zero, and enlarging the size of heat bath
can suppress the BE and accelerate the decay.

\subsection{Second bound entangled state}

We choose another $3\times 3$ bound entangled state as the initial state of
the two qutrits which was introduced by Bennett et al. \cite{Bennett2} from
the unextendible product bases:
\begin{eqnarray}
|\phi _{0}\rangle &=&\frac{1}{\sqrt{2}}|0\rangle (|0\rangle -|1\rangle ),\
\notag \\
\ |\phi _{1}\rangle &=&\frac{1}{\sqrt{2}}(|0\rangle -|1\rangle )|2\rangle ,
\notag \\
|\phi _{2}\rangle &=&\frac{1}{\sqrt{2}}|2\rangle (|1\rangle -|2\rangle ),\ \
\notag \\
|\phi _{3}\rangle &=&\frac{1}{\sqrt{2}}(|1\rangle )-|2\rangle )|0\rangle , \\
|\phi _{4}\rangle &=&\frac{1}{3}(|0\rangle +|1\rangle +|2\rangle )(|0\rangle
+|1\rangle +|2\rangle ),  \notag
\end{eqnarray}%
from which the density matrix could be expressed as
\begin{equation}
\rho =\frac{1}{4}(I_{9\times 9}-\sum_{j=0}^{4}|\phi _{j}\rangle \langle \phi
_{j}|),  \label{BE2}
\end{equation}%
In this case, the dynamics of entanglement are determined by six decoherence
factors, and analytical results are not available. We numerically calculate
BE and FE, and the results are shown in Fig.~5.
\begin{figure}[tbp]
\includegraphics[bb=13 179 581 656, width=8 cm, clip]{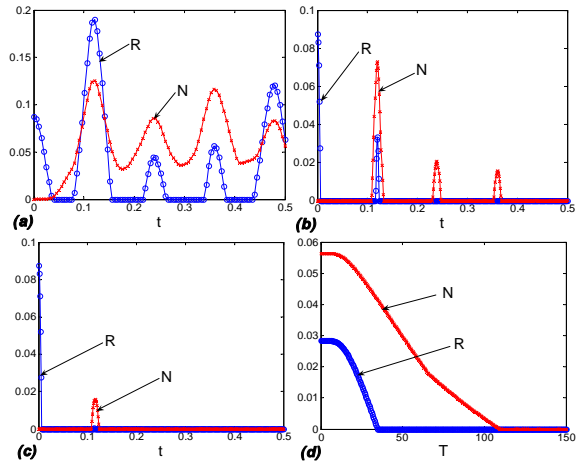}
\caption{We consider the BE and FE together. And in the four subfigures the
blue lines with circle markers correspond to $\mathcal{R}$ and the red lines
with $\times $ markers denote $\mathcal{N}$. \textbf{(a)} $\mathcal{R}$ and $%
\mathcal{N}$ versus time with coupling $g=1$, the size of environment $L=300$%
, and the system is at temperature $T=10$. The frequencies of heat bath are
chosen randomly in the region $\protect\omega _{j}\in \lbrack 50,55]$.
\textbf{(b)} Only changing the coupling to a larger one $g=5$, and the other
parameters are the same as subfigure \textbf{(a)}. In the figure \textbf{(c)}%
, parameter $\protect\delta =9$, the other parameters are the same as
subfigure \textbf{(b)}. In subfigure \textbf{(d)}, it shows $\mathcal{R}$
versus temperature at $t=0.005$ and $\mathcal{N}$ versus temperature at $%
t=0.115$. $g=1$ and $L=300$.}
\end{figure}

In order to compare BE and FE, we have numerically given the time behaviors
of both quantity $\mathcal{R}$ and negativity $\mathcal{N}$. We choose a
higher frequency region $\omega_j\in[50,50+\delta]$, which will induce some
interesting properties of BE and FE. In Fig.~5 {(a)}, we see that the
negativity can be nonzero, in contrast with the first BE given by Horodecki
et al. The nonzero negativity implies that the state is free entangled. The
state keeps bound entangled for a short time, and then, the state is free
entangled. When the coupling $g$ becomes stronger, as shown in Fig.~5 (b),
after some revivals, both quantities become zero.

In Fig.~5 {(c)}, we extend the width of frequency region to $\delta=9$.
Expanding the energy region makes the particles in heat bath own more
chances to take different energy. And this will prevent the revival of
entanglement. In Fig.~5{(d)}, we show the behaviors of $\mathcal{R}$ and $%
\mathcal{N}$ against temperature for a fixed time. With the increase of
temperature, quantity $\mathcal{R}$ and negativity $\mathcal{N}$ decrease
gradually, and at last the thermal fluctuation destroys the entanglement
completely.

\section{Bound entanglement in a spin environment}

To show the universality of the influence of environment on the time
evolution of BE, we need to use difference modeling of environment.
Here, we consider an environment consisting of $N$ spin halves. The
corresponding
model Hamiltonian reads~\cite{Zurek}%
\begin{equation}
H=\frac{g}{2}\left( S_{1z}+S_{2z}\right) \otimes \sum\limits_{k=1}^{L}\omega
_{k}\sigma _{z}^{(k)},  \label{Hspin}
\end{equation}%
where $\sigma _{z}^{(k)}$ denotes the $z$-component of the Pauli vector, and
$g$ denotes the coupling strength between central spins and environment. We
notice that the above model has been considered by Zurek \cite{Zurek} as a
solvable model of decoherence.

The time evolution operator can be expressed as:
\begin{equation}
U\left( t\right) =\prod\limits_{k=1}^{L}\exp (-it\hat{\Lambda}\omega
_{k}\sigma _{z}^{(k)}),
\end{equation}%
where we define a special operator-valued parameter
\begin{equation}
\hat{\Lambda}=\frac{g}{2}\left( S_{1z}+S_{2z}\right) .
\end{equation}

\subsection{Horodecki's Bound entangled state}

In a similar vein as the discussions of entanglement in the bosonic
environment, we first study the disentanglement of Horodecki's bound
entangled state and give the analytical results.

\subsubsection{Analytical results}

Let us consider the whole system initially starts from a product state
\begin{equation*}
\rho _{tot}\left( 0\right) =\rho _{a}\otimes \rho _{E},
\end{equation*}%
where the initial state of the two qutrits $\rho _{a}$\ is a mixed BE\ state
represented in Eq.~(\ref{BE1}), and $\rho _{E}$ denotes the initial state of
the environment which is assumed to be a thermal state described by the
density matrix
\begin{equation}
\rho _{E}=\prod\limits_{k=1}^{L}\frac{e^{\beta \omega _{k}\sigma _{z}^{(k)}}%
}{2\cosh (\beta \omega _{k})}.
\end{equation}%
Then the reduce density matrix at time $t$ can be given by the same matrix
as Eq.~(\ref{rho12}), and now the three decohrence factors in this spin
environment can be obtained as%
\begin{eqnarray}
\left\vert F_{1}\left( t\right) \right\vert &=&\left\vert F_{2}\left(
t\right) \right\vert =\prod\limits_{k=1}^{L}\left\vert F_{1,k}\right\vert
\notag \\
&=&\prod\limits_{k=1}^{L}\sqrt{1-\frac{\sin ^{2}(gt\omega _{k})}{\cosh
^{2}(\beta \omega _{k})}},  \label{r1} \\
\left\vert F_{3}\left( t\right) \right\vert
&=&\prod\limits_{k=1}^{L}\left\vert F_{3,k}\right\vert  \notag \\
&=&\prod\limits_{k=1}^{N}\sqrt{1-\frac{\sin ^{2}(2gt\omega _{k})}{\cosh
^{2}(\beta \omega _{k})}}.  \label{r2}
\end{eqnarray}

From the Eqs.~(\ref{r1}) and (\ref{r2}) one can find each
decoherence factor $\left\vert F_{k}\right\vert $ is less than
unity, which implies that in the large $L$\ limit, $\left\vert
F_{k}\left( t\right) \right\vert $ will go to zero under some
reasonable condition. Now, we make some further analysis by
introducing a cutoff number $K_{c}$\ similar to the discussion in
Ref.~\cite{quan}. We define the partial product as
\begin{equation}
\left\vert F_{1}\left( t\right) \right\vert
_{c}=\prod_{k>0}^{K_{c}}\left\vert F_{1,k}\right\vert \geq \left\vert
F_{1}\left( t\right) \right\vert ,  \label{cutoff}
\end{equation}%
from which the corresponding\ partial sum $\ln \left\vert F_{1}\left(
t\right) \right\vert _{c}\equiv -\sum_{k>0}^{K_{c}}\left\vert \ln
F_{1,k}\right\vert $. We can do some heuristic analysis in some special
conditions such as confining the energy spectrum in a region from zero to a
nonzero value $\omega _{k}\in \left[ 0,\omega \right] $. When the cutoff
number $K_{c}$ is small enough, in a finite long time, with some proper $g$
we can pick out some tiny $\omega _{k}$ to make $gt\omega _{k}$ begin a
small one and achieve the approximation $\sin ^{2}\left( gt\omega
_{k}\right) \approx (gt\omega _{k})^{2}.$ At a finite temperature we can
have
\begin{eqnarray}
\ln \left\vert F_{1}\left( t\right) \right\vert _{c} &=&\frac{1}{2}%
\sum_{k>0}^{K_{c}}\ln \left( 1-\frac{\sin ^{2}\left( gt\omega _{k}\right) }{%
\cosh ^{2}\left( \beta \omega _{k}\right) }\right)  \notag \\
&\approx &-\frac{1}{2}\left( \sum_{k>0}^{K_{c}}\frac{\omega _{k}^{2}}{\cosh
^{2}\left( \beta \omega _{k}\right) }\right) g^{2}t^{2}  \notag \\
&=&-\gamma t^{2}  \label{sum}
\end{eqnarray}%
where%
\begin{equation*}
\gamma =\frac{1}{2}g^{2}\sum_{k}^{K_{c}}\frac{\omega _{k}^{2}}{\cosh
^{2}\left( \beta \omega _{k}\right) }.
\end{equation*}

From Eqs.~(\ref{cutoff}) and (\ref{sum}), we find that the decoherence
factors decay in a Gaussian form with time, therefore from Eq.~(\ref%
{realignment}) it is apparent that the witness quantity $\mathcal{R}$ will
vanish in a finite time. Also from (\ref{sum}), if the temperature is very
low and quite nearly to $T=0$, $\gamma $ approaches zero, and the
decoherence factors quantity $\mathcal{R}$ will not decay with time. It
implies that, in our system, the temperature greatly affect the dynamics of
BE. It is a rough calculation in our analysis, nevertheless it gives us some
constructive results.

\subsubsection{Numerical results}

\begin{figure}[tbp]
\includegraphics[bb=20 300 572 539, width=8 cm, clip]{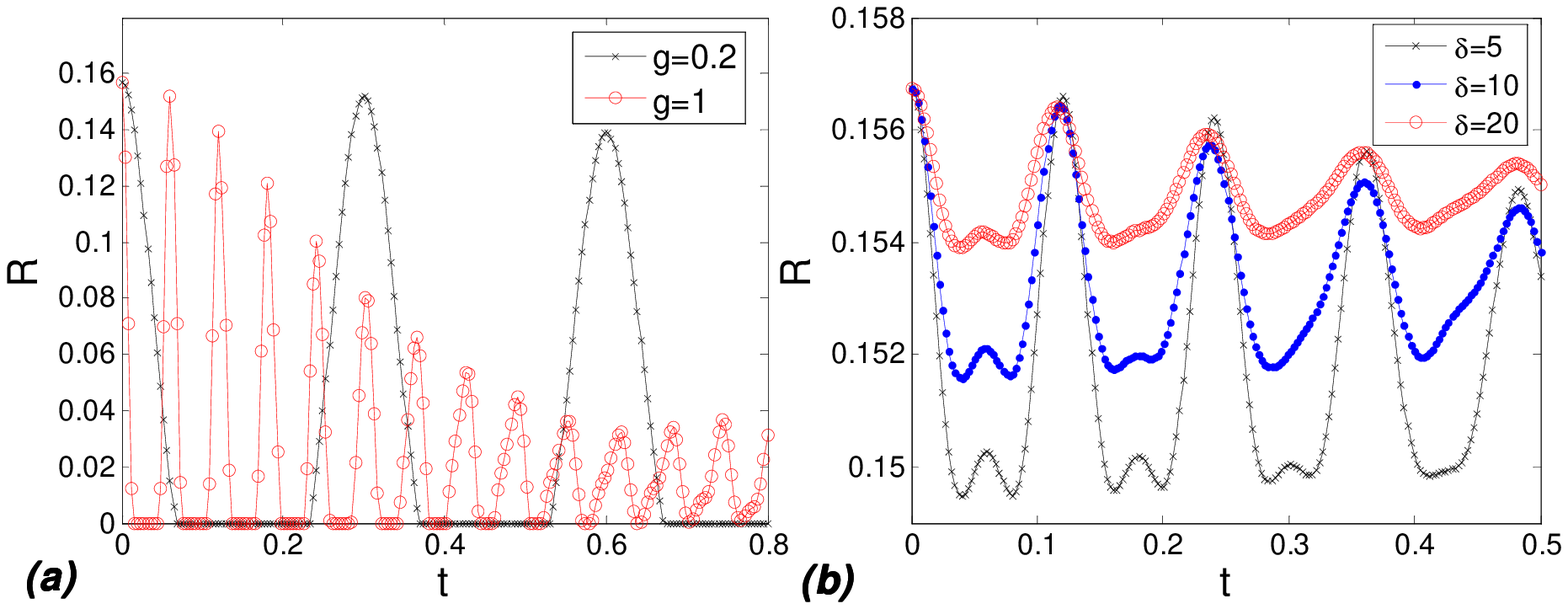}
\caption{\textbf{(a)} $\mathcal{R}$ versus time with different coupling
parameter $g$. The couplings $\protect\omega_k$ of the spins in environment
is random in the region $\protect\omega_k\in[50, 55]$. The size of the spin
environment is $L=300$ and the system is at the temperature $T=15$. \textbf{%
(b)} shows $\mathcal{R}$ versus time with different $\protect\delta$ which
is the region widths of the couplings $\protect\omega_k$, and $\protect\omega%
_k\in[50,50+\protect\delta]$. }
\end{figure}
\begin{figure}[tbp]
\includegraphics[bb=17 212 566 645, width=8 cm, clip]{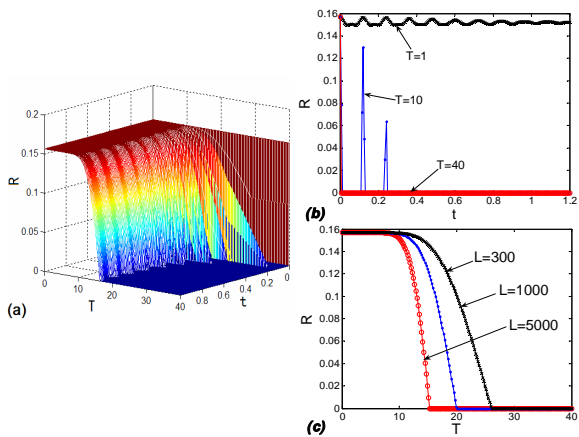}
\caption{\textbf{(a)} Three dimensional (3D) diagram of the $\mathcal{R}$
versus time and temperature, with $L=300$, $g=0.5$ and $\protect\omega_k \in[%
50, 55]$. \textbf{(b)} Several part sections of the 3D diagram. Explicitly,
it presents $\mathcal{R}$ versus time at different temperatures of $T=1$, $%
T=10$ and $T=40$. \textbf{(c)} shows $\mathcal{R}$ decays with temperature
at a fixed time $t=0.005$ with different sizes of environment $L=300, 1000,$
and $5000$.}
\end{figure}
If the frequency distribution are in the region $\omega_k\in[0,\omega]$, the
decoherence factors displays a Gaussian decay, which is analytically studied
in the former section. And in Ref.~\cite{Cucchietti}, Gaussian decay was
shown numerically in various distributions of couplings. So, here, we
consider the distributions such as $[50,50+\delta]$, a higher frequency
region. In Fig.~6 {(a)}, the larger coupling strength $g$ makes a stronger
oscillations of BE, which is different from the case of bosonic environment.
Mathematically, we can understand that from Eq.~(\ref{r1}) and (\ref{r2}),
parameter $g$ can change the frequency of the periodic function. The
collapses and revivals of $\mathcal{R}$ is also observed here, and the
revival amplitude decreases with time. Fig.~6 {(b)} is a plot of $\mathcal{R}
$ for different width of frequency distribution. The wider distribution will
smear the collapse and revival phenomenon, namely, the BE evolve smoothly
with time, and the BE is always there.

Effects of the thermal fluctuation on the dynamic of BE are shown in Fig.~7.
The 3D plot {(a)} displays a flat at low temperatures about $T<10$, which
implies that the BE is stable against temperature in this region. When
temperature is high enough, BE rapidly decreases to zero. In Fig.~7{(b)}, we
can see explicitly that temperature plays an important role in the dynamics
of BE in this spin environment. It is just like a control process that when
temperature is higher than some value, BE will decay sharply with time. At a
very low temperature, from Eq.~(\ref{r1}) and Eq.~(\ref{r2}), we know that
decoherence factors are approximately one with very small oscillations. And
the BE is of a little change with time, namely, the BE is stable against
time for low temperatures. This is a quite different property from the case
of bosonic environment. In Fig.~7 {(c)}, we also plot the BE against
temperature for different size of environment. As we expected, the larger
size of the spin environment accelerates the decaying process.

\subsection{Second bound entangled state}

\begin{figure}[tbp]
\includegraphics[bb=74 240 496 593, width=6 cm, clip]{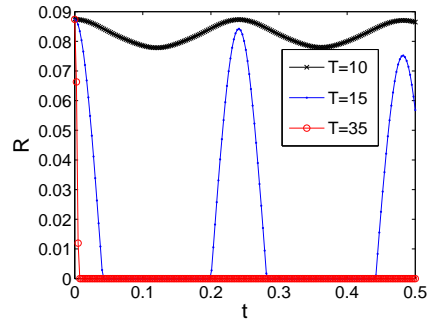}
\caption{$\mathcal{R}$ versus time at different temperatures, $T=10$, $T=15$
and $T=35$. With the coupling $g=0.5$, environment size $L=300$ and $\protect%
\omega_k\in[50,55]$.}
\end{figure}
We now consider the case that the two qutrits initially in the second bound
entangled state (\ref{BE2}) in the spin environment. We choose $\omega_k\in[%
50,55]$ and study the finite-temperature effects on entanglement. In Fig.~8,
we can see the similar phenomena with Fig.~6, namely, at low temperature BE
only oscillates around the initial value of the BE. At higher temperature BE
decays sharply with time, and revivals are also observed with revival
amplitudes decreasing with time. We also study negativity, however, it
always keeps zero which means there does not exist FE all the time.

\section{\protect\vspace{0pt}conclusion}

In summary, we have studied the decoherence phenomena of two-qutrit
system couple to an enviroment when this open systems are initially
prepared in a bound entangled state. The so-called "sudden death"
phenomenon~\cite{Yu} of entanglement can be found in our studies as
a common feature of BE evolution. Two typical pure dephasing
systems, the bosonic and the spin systems are considered in order to
show this kind of universality of decoherence process with BE. Here,
we used the realignment criterion to characterize BE, and the PPT
criterion to study FE. Beyond the negativity, we have introduced a
novel witness quantity $\mathcal{R}$ to study BE. Those two
approaches are operational and convenient to use. Two kinds of BE in
initial state of the open system are considered, one is given by
Horodecki et al, and another is constructed from the unextendible
product basis.

One of our central result is to express the quantity $\mathcal{R}$ and
negativity $\mathcal{N}$ in terms of three decoherence factors, and these
factors are analytically obtained. In the case of bosonic environments, the
Gaussian decay and the exponential decay of the BE was found, and in the
case of spin environment, we find that BE can display a Gaussian decay. In
both environments, the collapse and revivals of the BE are observed for
system frequency distribution being in a higher region with an appropriate
width. Larger coupling strength $g$, larger environment size $L$ and higher
temperature will enhance the disentanglement process. For the Horocecki's BE
in the spin environment, we find that the BE can be stable against low
temperature increase.

Finally we have to point out that since we are lack of entanglement measure
for two qutrits, the study of decoherence of entanglement here is
incomplete. Nevertheless, the realignment criterion and the PPT criterion
are very efficient to characterize BE. It will be interesting to consider
decoherence of BE under other decoherence processes such as dissipation, and
investigate the robustness of the BE.

\acknowledgements This work is supported by NSFC with No. 10405019,
10604002, 10474104, 90503003, 60433050, the specialized Research
Fund for the Doctoral Program of Higher Education (SRFDP) under
grant No.20050335087, and the National Basic Research Program (also
called 973 Program) under grant No.2006CB921206, 2005CB724508. We \
thanks H. Dong, T. Shi and L. Zhou for some valuable discussions.

\end{document}